\documentclass[aps,prd,onecolumn,showpacs,groupedaddress,nofootinbib]{revtex4-2}
\usepackage{graphicx}
\usepackage{dcolumn}
\usepackage{bm}
\usepackage{color}
\usepackage{longtable}
\usepackage{supertabular}
\usepackage[T1]{fontenc}
\usepackage{epstopdf}
\usepackage{amsmath}
\usepackage{amssymb}
\usepackage{setspace}
\usepackage{multirow}
\usepackage{booktabs}
\usepackage[section]{placeins}
\usepackage{mathrsfs}
\usepackage{hyperref}

\makeatletter

\newcommand{\Rmnum}[1]{\expandafter\@slowromancap\romannumeral #1@} 
\newcommand{\bq}{\begin{equation}}
\newcommand{\eq}{\end{equation}}
\newcommand{\bqn}{\begin{eqnarray}}
\newcommand{\eqn}{\end{eqnarray}}
\newcommand{\nb}{\nonumber}
\newcommand{\lb}{\label}

\makeatother

\begin{document}
\title{Late-time tail and echoes of Damour-Solodukhin wormholes}

\author{Wei-Liang Qian\textsuperscript{2,1,3}}\email[E-mail: ]{wlqian@usp.br (corresponding author)}
\author{Qiyuan Pan\textsuperscript{4,1}}
\author{Bean Wang\textsuperscript{5}}
\author{Rui-Hong Yue\textsuperscript{1}}

\affiliation{$^{1}$ Center for Gravitation and Cosmology, College of Physical Science and Technology, Yangzhou University, Yangzhou 225009, China}
\affiliation{$^{2}$ Escola de Engenharia de Lorena, Universidade de S\~ao Paulo, 12602-810, Lorena, SP, Brazil}
\affiliation{$^{3}$ Faculdade de Engenharia de Guaratinguet\'a, Universidade Estadual Paulista, 12516-410, Guaratinguet\'a, SP, Brazil}
\affiliation{$^{4}$ Key Laboratory of Low Dimensional Quantum Structures and Quantum Control of Ministry of Education, Synergetic Innovation Center for Quantum Effects and Applications, and Department of Physics, Hunan Normal University, Changsha, Hunan 410081, China}
\affiliation{$^{5}$ Department of Physical Sciences and Applied Mathematics, Vanguard University, Costa Mesa, CA 92626, USA}

\begin{abstract}
Damour-Solodukhin wormholes are intriguing theoretical constructs, closely mimicking many properties of black holes. 
This study delves into two distinct characteristics of the waveforms emitted from such wormholes, namely, the late-time tails and echoes, which can substantially be used to distinguish its identity. 
Notably, both features appear in the latter stages of quasinormal oscillations and stem from the singularities of the Green's function. 
The late-time tail, on the one hand, arises due to the branch cuts in the relevant Green's function. 
Within the Damour-Solodukhin wormhole paradigm, singularities are present in both ingoing and outgoing waveforms, which entails a generalization of the existing recipe for black hole metrics.
On the other hand, the echoes are attributed to a new set of quasinormal poles, supplementing those of the respective black holes, reminiscent of the scenario where the spacetime metric possesses a discontinuity. 
It is inferred that both features are observationally relevant in distinguishing a wormhole from its black hole counterpart. 
Moreover, we suggest a potential interplay concerning the late-time evolution between the two mechanisms in question. 
\end{abstract}

\date{June 16th, 2024}

\maketitle

\newpage
\section{Introduction}\label{sec1}

As one of the most captivating concepts in theoretical physics, the black hole demonstrates the properties of gravity at its extremity. 
Successful detections of gravitational waves emanated from the binary mergers achieved by the ground-based LIGO and Virgo collaboration~\cite{agr-LIGO-01, agr-LIGO-02, agr-LIGO-03, agr-LIGO-04} established the inauguration of a new era for observational astrophysics. 
The accomplishment has further inspired analysis of ring down waveforms~\cite{agr-BH-spectroscopy-15, agr-BH-spectroscopy-16, agr-BH-spectroscopy-35, agr-BH-spectroscopy-38, agr-BH-spectroscopy-39}.
Moreover, it has been speculated~\cite{agr-LISA-19, agr-LISA-20, agr-TianQin-05} that the ringdown signals will be more pronounced regarding the ongoing space-borne projects LISA~\cite{agr-LISA-01}, TianQin~\cite{agr-TianQin-01, agr-TianQin-Taiji-review-01}, and Taiji~\cite{agr-Taiji-01}.

Such specific waveforms primarily consist of quasinormal modes (QNMs)~\cite{agr-qnm-review-02, agr-qnm-review-03, agr-qnm-review-06}.
The physical interest arises from the fact that they are uniquely dictated by the underlying spacetime, and as a result, their observation furnishes unambiguous information on the spacetime's properties in the vicinity of the horizon.
It was pointed out by Leaver~\cite{agr-qnm-21, agr-qnm-29} that the QNMs are associated with the poles pertaining to Green's function of the underlying master equation, while the late-time tail is attributed to the branch cut, which usually follows a form that drops off as an inverse power in time~\cite{agr-qnm-tail-01}.
For massless perturbations, the branch cut is typically placed on the negative imaginary axis in the frequency domain~\cite{agr-qnm-tail-05, agr-qnm-tail-06}, while for massive ones, it can be placed on the real axis between two branch points~\cite{agr-qnm-tail-12, agr-qnm-tail-21, agr-qnm-tail-39}.
The appearance of the late-time tail also indicates that the QNM poles are incomplete regarding the description of the waveforms.

More recently, Cardoso {\it et al.} suggested the possibility of echoes~\cite{agr-qnm-echoes-01, agr-qnm-echoes-02}, another intriguing concept that intervened with the late-stage ringing waveforms. 
As a potential observable that might tell apart different but otherwise similar gravitational systems via their distinct properties near the horizon, the idea has incited many studies of echoes in various systems.
The ballpark of the latter embraces exotic compact objects like gravastar~\cite{agr-eco-gravastar-02, agr-eco-gravastar-03}, wormhole~\cite{agr-wormhole-01, agr-wormhole-02, agr-wormhole-10, agr-wormhole-11}, among others~\cite{agr-qnm-echoes-29, agr-qnm-echoes-30}.
In particular, a Damour-Solodukhin type wormhole~\cite{agr-wormhole-12, agr-wormhole-13} closely mimics a static black hole in terms of apparently irreversible accretion of matter, no-hair properties, and quasinormal ringing.
In this regard, echoes are one of the promising features that can be used to distinguish its identity from its counterpart.

Similar to the late-time tail, echoes can also be attributed to the analytic properties of the master equation, and in particular, its Green's function.
Mark {\it et al.}~\cite{agr-qnm-echoes-15} evaluated Green's function in the frequency domain.
Subsequently, echoes in compact objects are derived by rewriting the response waveform as a summation of a geometric series of products in reflection and transmission amplitudes while evaluating the convolution integration involved in the inverse Fourier transform.
This approach was borrowed to explore the scenario where the metric possesses a discontinuity by some of us~\cite{agr-qnm-echoes-20}.
For the latter, the emergence of echoes is understood in terms of the asymptotical pole structure of QNM spectrum~\cite{agr-qnm-lq-03, agr-qnm-lq-matrix-06}, and consequently, the discontinuity gives rise to an alternative mechanism for echoes.
Echoes in the context of Damour-Solodukhin type wormholes~\cite{agr-wormhole-12} have been explored by Bueno {\it et al.}~\cite{agr-qnm-echoes-16}.
By explicitly solving for specific frequencies when the transition matrix becomes singular, the obtained quasinormal frequencies lie uniformly along the real axis, whose interval manifestly leads to echoes.

From an empirical perspective, while the two concepts are discussed in the literature mainly in a separate fashion, both the tail and echoes appear in the late stage of the evolution.
They are also related from an analytic viewpoint, as both scenarios are closely associated with the singularities of the underlying Green's function.
On the one hand, for the late-time tail, the fact that the branch cut stems from or lies on the real axis guarantees that the phenomenon lends a significant contribution in the late stage of the temporal evolution.
On the other hand, for echoes, the proximity between the asymptotic quasinormal spectrum and the real axis ensures that the phenomenon manifests at the evolution's late stage. 
The present study is motivated by the above consideration to elaborate the two phenomena on a consistent footing.
Specifically, we perform our analysis in the context of Damour-Solodukhin wormhole metrics.
For the late-time tail, we generalize the treatment in the seminal paper by Ching {\it et al.}~\cite{agr-qnm-tail-05, agr-qnm-tail-06} to the cases where both the ingoing and outgoing waveforms possess singularities.
As it turns out, such a procedure is not straightforward as directly flipping both the ingoing and outgoing waveform to the other side of the branch cut leads to a trivial power law form.
We delve into some subtleties that involve the branch cuts in contour integrations and show that proper evaluation of the Wronskian implies a normalization of the waveform. 
On the other hand, for the echoes in the gravitational system in question, we show that they are attributed to a collective effect of a novel branch of quasinormal poles, where the quasinormal spectrum of the original black hole largely remains unchanged. 
As the interval along the real axis between successive poles is typically smaller than the real part of the fundamental mode, the echoes modulate the envelope of the original ringdown signals of the black hole metric.
Besides, we explicitly show the equivalency between the two approaches given in Refs.~\cite{agr-qnm-echoes-15} and~\cite{agr-qnm-echoes-16}.
Moreover, based on these discussions, we suggest an interplay between the two mechanisms concerning the late-time evolution.
Specifically, by numerical simulations, we demonstrate the possibility of the above where one may predominate. 

The remainder of this paper is organized as follows. 
In Sec.~\ref{sec2}, we briefly review Green's function approach in black hole perturbation theory.
Subsequently, in Sec.~\ref{sec3}, the analytic recipe of late-time tail in black hole metric is revisited and generalized to the case where both the ingoing and outgoing waveforms contain branch cuts.
A derivation for echoes in Damour-Solodukhin type wormholes is presented in Sec.~\ref{sec4}, initially proposed for the black hole metric with discontinuity.
As a collective effect from a particular branch of asymptotic QNMs, echoes modulate the quasinormal oscillations in the late stage.
Furthermore, in Sec.~\ref{sec5}, we argue the possibility of an interplay between these two late-time phenomena in the context of Damour-Solodukhin-type wormholes.
Such a scenario is illustrated numerically for both the approximated model and the Schwarzschild Damour-Solodukhin wormhole.
The concluding remarks are given in the last section.
The complementary mathematical derivations will be relegated to Appx.~\ref{appA},~\ref{appB}, and~\ref{appC}.

\section{Green's function approach for black hole perturbations}\label{sec2}

In this section, we briefly revisit Green's function formalism employed in the black hole perturbation theory, which furnishes the basis for the analysis carried out in the preceding sections.

In practice, the study of black hole perturbation theory often leads to exploring the solution of the radial part of the master equation~\cite{agr-qnm-review-03,agr-qnm-review-06},
\begin{eqnarray}
\frac{\partial^2}{\partial t^2}\Psi(t, x)+\left(-\frac{\partial^2}{\partial x^2}+V_\mathrm{eff}\right)\Psi(t, x)=0 ,
\label{master_eq_ns}
\end{eqnarray}
where the spatial coordinate $x$ is known as the tortoise coordinate, and the effective potential $V_\mathrm{eff}$ is governed by the given spacetime metric, spin ${\bar{s}}$, and angular momentum $\ell$ of the waveform.
For instance, it can be the Regge-Wheeler potential $V_\mathrm{RW}$ for the Schwarzschild black hole metric
\bqn
V_\mathrm{eff} = V_\mathrm{RW}=F\left[\frac{\ell(\ell+1)}{r^2}+(1-{\bar{s}}^2)\frac{r_h}{r^3}\right],
\lb{Veff_RW}
\eqn
where 
\bqn
F=1-r_h/r ,
\lb{f_RW}
\eqn
where the horizon $r_h=2M$ and $M$ is the mass of the black hole, and the tortoise coordinate is related to the radial coordinate $r$ by the relation $x=\int dr/F(r)$.

The black hole QNMs are determined by solving the eigenvalue problem defined by Eq.~\eqref{master_eq_ns} in the frequency domain
\begin{equation}
\frac{d^2\Psi(\omega, x)}{dx^2}+[\omega_n^2-V_\mathrm{eff}(r)]\Psi(\omega, x) = 0 , \label{eq2}
\end{equation}
per the following boundary conditions for asymptotically flat spacetimes
\begin{equation}
\Psi \sim
\begin{cases}
   e^{-i\omega_{n} x}, &  x \to -\infty, \\
   e^{+i\omega_{n} x}, &  x \to +\infty,
\end{cases}
\label{master_bc0}
\end{equation}
which indicates an ingoing wave at the horizon and an outgoing wave at infinity.
The subscript $n$ represents the overtone number.
Apart from the initial burst, the temporal profile of the waveform $\Psi$ is characterized by the quasinormal oscillations and late-time tail.
The QNMs are governed by the eigenvalues $\omega_{n}$, known as the quasinormal frequencies.
They are typically complex numbers attributed to the dissipative nature of Eq.~\eqref{master_bc0}.
On the other hand, the tail contribution is governed by the asymptotical properties of the effective potential, and subsequently, the physical content is referred to as the backscattering of perturbed wave packets by spacetime far away from the black hole~\cite{agr-qnm-tail-06, agr-qnm-tail-20, agr-qnm-tail-31, agr-qnm-tail-08, agr-qnm-tail-09, agr-qnm-tail-32}.

Both the QNMs and late-time tail are associated with the analytic properties of the underlying Green's function that satisfies
\begin{equation}
\left[\frac{d^2}{dx^2}+(\omega_{n}^2-V_\mathrm{eff}(r))\right]\widetilde{G}(\omega, x,y)= \delta(x-y) .\label{DefGreen}
\end{equation}
According to the standard procedure~\cite{agr-qnm-21, agr-qnm-28, agr-qnm-29}, Green's function can be constructed using the form
\begin{equation}
\widetilde{G}(\omega, x,y)= \frac{1}{W(\omega)}f(\omega, x_<)g(\omega, x_>) ,\label{FormalGreen}
\end{equation}
where $x_<\equiv \min(x, y)$, $x_>\equiv \max(x, y)$, and
\begin{equation}
W(\omega) \equiv W(g, f) = {g} {f}' - {f} {g}' \label{DefWronskian}
\end{equation}
is the Wronskian of $f$ and $g$, 
where $f$ and $g$ are the two linearly independent solutions of the corresponding homogeneous equation satisfying the boundary conditions Eq.~\eqref{master_bc0} at the horizon and infinity.
To be specific, $f$ and $g$ possess the following asymptotic forms
\begin{equation}
f(\omega, x) \sim
\begin{cases}
   e^{-i\omega x}, &  x \to -\infty, \\
   A_{\mathrm{out}}(\omega)e^{+i\omega x}+A_{\mathrm{in}}(\omega)e^{-i\omega x}, &  x \to +\infty,
\end{cases}
\label{master_bc1}
\end{equation}
and
\begin{equation}
g(\omega, x) \sim
\begin{cases}
   B_{\mathrm{out}}(\omega)e^{+i\omega x}+B_{\mathrm{in}}(\omega)e^{-i\omega x}, &  x \to -\infty,\\
   e^{+i\omega x}, &  x \to +\infty, 
\end{cases}
\label{master_bc2}
\end{equation}
in asymptotically flat spacetimes, which are bounded at the limit $t\to +\infty$ for $\Im \omega <0$.
In the above expressions, $A_{\mathrm{in}}$, $A_{\mathrm{out}}$, $B_{\mathrm{in}}$, and $B_{\mathrm{out}}$ are the reflection and transmission amplitudes, whose specific forms might be unknown to us but are well-defined for a given metric in principle.
It is noted that the amplitudes of the waveforms satisfy the relations~\cite{book-blackhole-Frolov}
\bqn
B_{\mathrm{out}} &=& A_{\mathrm{in}} ,\nb\\
B_{\mathrm{in}} &=& -A_{\mathrm{out}}^* ,\label{fluxConv}
\eqn
as a result of completeness and flux conservation.
Besides, the black hole's reflection and transmission amplitudes are defined as
\bqn
\widetilde{\mathcal{R}}_\mathrm{BH}(\omega) &=& \frac{B_{\mathrm{in}}}{B_{\mathrm{out}}},\nb\\
\widetilde{\mathcal{T}}_\mathrm{BH}(\omega) &=& \frac{1}{B_{\mathrm{out}}} ,\label{RefTransA}
\eqn
for an outgoing wave coming from $-\infty$.

The QNMs correspond to the pole of Green's function.
While the intrinsic pole structure primarily comes from the zeros of the Wronskian Eq.~\eqref{DefWronskian}, it is worth noting that it might suffer modifications owing to mechanisms such as pole skipping~\cite{adscft-pole-skipping-02, adscft-pole-skipping-05} and external source might also bring in additional distortion~\cite{agr-qnm-lq-02}.

\section{Late-time tail in Damour-Solodukhin type wormholes}\label{sec3}

In what follows, we first outline the strategy to evaluate the asymptotical properties of late-time tail, closely following Refs.~\cite{agr-qnm-tail-05, agr-qnm-tail-06}.
The formalism is then generalized to the Damour-Solodukhin-type wormholes.

\subsection{The standard formalism to determine the power-law in late-time tail}

The late-time tail is primarily governed by the properties of effective potential at the asymptotic spatial infinity.
Analytically, it becomes apparent as one performs the inverse Fourier transform to attain the time-domain waveform by integrating along the branch cut and receives most contributions from the vicinity of the real frequency axis.
On the numerical side, this can be demonstrated by truncating the effective potential at an arbitrary but spatially finite location.
For the latter case, the tail phenomenon disappears since the effective potential vanishes sufficiently fast.

In practice, the Regge-Wheeler potential Eq.~\eqref{Veff_RW} does not imply explicit analytic forms for $f$ and $g$.
However, as the late-time tail is reflection of the asymptotic form of the metric at spatial infinity, it suffices to consider the corresponding asymptotic solutions~\cite{agr-qnm-tail-06, agr-qnm-tail-12, agr-qnm-tail-13, agr-qnm-tail-20, agr-qnm-tail-21, agr-qnm-tail-22, agr-qnm-tail-31, agr-qnm-tail-32, agr-qnm-tail-33}.
On the other hand, near the horizon, it can be argued that for most generic black hole metrics, the effective potential vanishes sufficiently fast near the horizon~\cite{agr-qnm-tail-39}.
Precisely, the effective potential must vanish asymptotically faster than an exponential form, and subsequently, $f(\omega, x)$ does not contain any singularity.
In other words, the specific waveform is irrelevant apart from its asymptotic form given by the first line of Eq.~\eqref{master_bc1}.
As a result, one only needs to deal with the singularity embedded in $g(\omega, x)$.

Based on the above arguments, regarding the outgoing wave, one may effectively ignore the factor $F$ in Eq.~\eqref{Veff_RW} as it satisfies $F\to 1$ as $x\to \infty$ and assume that the effective potential is significantly suppressed as $x\to -\infty$. 
Specifically, one considers the following form
\begin{equation}
V^\mathrm{BH}_\mathrm{eff}(x,\nu) = 
\begin{cases}
    \frac{\nu(\nu+1)}{x^2} + \delta V_\mathrm{eff} &\ \ \ \mathrm{region}\ 1\ \ \  x \ge x_c,\\
    \mathrm{suppressed} &\ \ \ \mathrm{region}\ 2\ \ \   x < x_c, 
\end{cases}  \label{Veff_alpha}
\end{equation}
where $\nu$ is a constant but not an integer and
\begin{equation}
\delta V_\mathrm{eff} = (1-{\bar{s}}^2)\frac{r_h}{x^3} , \label{Veff_2}
\end{equation}
for $x \ge x_c > 0$, where $x_c$ is a generic spatial point.

As $x\to+\infty$, it is apparent that the outgoing waveform is primarily governed by the first term on the r.h.s. of Eq.~\eqref{Veff_alpha}.
Moreover, further analysis~\cite{agr-qnm-tail-06} reveals that the contribution to the tail phenomenon is also entirely determined by the approximate waveform's singularity as long as $\nu$ is not an integer (c.f. $\ell$ in the Regge-Wheeler potential). 
Therefore, for simplicity, in the present study, we will focus on the case where one ignores the correction $\delta V_\mathrm{eff}$ where the outgoing wave possesses the form
\begin{equation}
g(\omega, x)=e^{i(\rho+1)\frac{\pi}{2}}\sqrt{\frac{\pi\omega x}{2}}H_\rho^{(1)}(\omega x) , \label{gH2}
\end{equation}
where $\rho=\nu+\frac12$ and $H_\rho^{(1,2)}$ is the Hankel function of the first (or second) kind~\cite{book-methods-mathematical-physics-05} so that appropriate boundary condition is attained asymptotically
\begin{equation}
g(\omega, x)\sim e^{i\omega x}, \ \ \mathrm{as} \ \ x\to +\infty \ \ \mathrm{for} \ \ -\pi<\mathrm{arg}\omega <2\pi  . \label{gH2asym}
\end{equation}

When viewed as a series expansion, it is apparent that $H_\rho^{(1)}$ contains a branch point at $\omega=0$, and therefore, the analytic extension to the complex frequency plane is not unique.
Nonetheless, it can be argued that the corresponding branch cut must be placed below the real frequency axis. 
In other words, given that the asymptotic boundary condition Eq.~\eqref{gH2asym} must be satisfied for any arbitrary real frequency $\omega$, all feasible options to analytically extend the domain of the frequency into the complex plane inevitably places the branch cut below the real axis.

For instance, the following analytic extension 
\begin{equation}
-\frac12\pi \le \mathrm{arg}\omega = \mathrm{arg}(\omega x) \le \frac43\pi  \label{anaCon1}
\end{equation}
is a possible choice as it satisfies the constraint implied by Eq.~\eqref{gH2asym}, which subsequently places a branch cut on the negative imaginary axis.
In contrast, it is not possible to perform an analytic extension for which the branch cut sits above the real axis, while the asymptotic waveforms possess the form given by Eq.~\eqref{master_bc2} as $\omega\to \pm\infty$.

The properties of the late-time tail can be analyzed by deriving the time-domain Green's function, which is obtained by the inverse Fourier transform
\begin{equation}
G(t,x,y) =\int_{-\infty}^{\infty} \frac{d\omega}{2\pi} \widetilde{G}(\omega, x, y) e^{-i\omega t}  . \label{invFourTrans}
\end{equation}
Based on Eq.~\eqref{FormalGreen} and the above discussions, one only needs to focus on the branch cut in $g(\omega, x)$.
According to Jordan's Lemma, one can enclose the contour integration in the clockwise direction via the large circle in the lower half of the frequency plane, which, in turn, naturally picks up the contribution from the branch cut placed on the negative imaginary axis according to Eq.~\eqref{anaCon1}.
On the contrary, if the branch cut is not below the real frequency axis, the late-time feature in the time-domain retarded Green's function will be absent.
Therefore, one concludes that the presence of such a feature dictates that the branch cut's placement cannot be arbitrary.
In this regard, one can further show that the choice Eq.~\eqref{anaCon1} is equivalent to those where the branch cut is below the real frequency axis.

The branch cut contributes to the integrant of Eq.~\eqref{invFourTrans} in terms of the difference of waveform $g(\omega, x)$ and the Wronskian $W(\omega)$ on the two sides of the branch cut.
Specifically, for $y \le x_c \le x$, we have
\bqn
\widetilde{G}_\pm(\omega, x,y) 
=\frac{f(\omega, y)g_\pm(\omega, x)}{W_\pm(\omega)} 
=\frac{f(\omega, y)g_\pm(\omega, x)}{W(g_\pm, f)}, \label{Gtilde_pm}
\eqn
where we have used ``$\pm$'' to denote the quantity on the right (left) side of the imaginary axis and $g_\pm(\omega, x)$ are found to attain the following forms~\cite{agr-qnm-tail-06}
\bqn
g_+(\omega=-i\sigma, x) &=& \sqrt{\frac{i\pi\sigma x}{2}}e^{\frac12 i\nu \nu\pi}\left[2\cos\rho\pi H_\rho^{(1)}(i\sigma x)+e^{-i\rho\pi} H_\rho^{(2)}(i\sigma x)\right], \nb\\
g_-(\omega=-i\sigma, x) &=& \sqrt{\frac{i\pi\sigma x}{2}}e^{\frac12 i\nu \nu\pi}\left[e^{-i\rho\pi} H_\rho^{(2)}(i\sigma x)\right] , \label{gH2pm}
\eqn
where $\sigma >0$.
In deriving the above results, one notes that the waveform Eq.~\eqref{gH2} takes a real value on the positive imaginary axis.
Subsequently, Eq.~\eqref{invFourTrans} can be rewritten as an integral on the negative imaginary axis
\bqn
G(t,x,y) &=& \int_{0}^{-i\infty} \frac{d\omega}{2\pi} \Delta\widetilde{G}(\omega, x, y) e^{-i\omega t}  \nb\\
&=& -i\int_{0}^{\infty} \frac{d\sigma}{2\pi} \left[G_+(-i\sigma, x, y)-G_-(-i\sigma, x, y)\right] e^{-\sigma t}  \nb\\
&=& -i\int_{0}^{\infty} \frac{d\sigma}{2\pi} \frac{W(g_-,g_+)}{W(g_+,f)W(g_-,f)}f(-i\sigma,x)f(-i\sigma,y) e^{-\sigma t}, \label{IntBranchCut}
\eqn
where on the branch cut $\omega=-i\sigma$.
Due to the suppression in the factor $e^{-\sigma t}$, the contribution of the integral comes primarily from the region $\sigma\to 0_+$.

The resulting asymptotic form of the late-time tail can be inferred from the frequency's dominant power factors of the waveforms $f(\omega)$, $g(\omega)$, and the Wronskian.
To be specific, evaluating the Wronskian and using the asymptotic expansion of the Hankel function, we have
\bqn
W(g_-,g_+) &=& -4i\sigma \sin\nu\pi ,\nb\\
g_\pm(-i\sigma, x)&\propto& \sigma^{-\nu} , \nb\\
W(g_\pm,f)&=& \left[f'(-i\sigma,x_c)+\frac{\nu}{x_c}f(-i\sigma,x_c)\right]g_\pm(x_c) \propto \sigma^{-\nu} ,\label{assForms}
\eqn
where $x_c$ is a rather arbitrary choice of spatial coordinate where the waveforms are evaluated.
For Eq.~\eqref{IntBranchCut} to be meaningful, $\nu$ cannot be an integer.
By taking into consideration that $f(\omega, x)$, when written as Taylor expansion in $\omega$, does not contain any peculiar properties, one finds
\bqn
\Delta\widetilde{G}(\omega, x, y) \propto  \sigma^{2\nu+1} ,
\eqn
and therefore Eq.~\eqref{IntBranchCut} gives the power law
\bqn
G(t,x,y) \propto \int_{0}^{\infty} d\sigma \sigma^{2\nu+1} e^{-i\omega t} \propto  t^{-(2\nu+2)} .
\eqn

The above derivation features a few assumptions that might not be valid for more sophisticated scenarios.
\begin{itemize}
    \item There is no singularity in the ingoing waveform $f(\omega, x)$.
    \item Even though the Wronskian is independent of the spatial coordinate, it is explicitly evaluated at a specific coordinate $x_c$ where the forms for both the waveforms $f(\omega, x)$ and $g(\omega, x)$ are accessible, namely, either approximately obtained or assumed.
\end{itemize}
Nonetheless, many fellow studies in the literature have widely adopted these assumptions.

\subsection{Generalization to the case of singular ingoing wave}

If there is some singularity in the ingoing waveform $f(\omega, x)$, it is apparent that the standard procedure elaborated in the last section must be modified.
In this subsection, we first argue that such a mathematical complication is physically relevant in the context of wormhole metrics.
Subsequently, we show that the treatment implies a proper normalization for the Wronskian when evaluating Green's function. 

The Damour-Solodukhin wormholes~\cite{agr-wormhole-12} are interesting foils whose properties are mainly reminiscent of their black hole counterparts.
As a result, features such as late-time tails and echoes are crucial in distinguishing these objects from the corresponding black holes.

The effective potential of a Damour-Solodukhin wormhole is typically mirrored with respect to the wormhole's throat.
While on each side of the throat, the potential is numerically close to the black hole effective potential outside the horizon.
In this regard, we consider the following symmetric effective potential:
\begin{equation}
V^\mathrm{DS}_\mathrm{eff}(x) = 
\begin{cases}
   V^\mathrm{BH}_\mathrm{eff}(x, \nu_1)   &\ \ \mathrm{region}\ 1 \ \ \ x > x_c, \\
   V_c   &\ \ \mathrm{region}\ 2 \ \ \ -x_c\le x \le x_c, \\
   V^\mathrm{BH}_\mathrm{eff}(-x,\nu_3)   &\ \ \mathrm{region}\ 3 \ \ \ x < -x_c, 
\end{cases}  \label{Veff_DS}
\end{equation}
where $x_c > 0$ is a generic coordinate indicating the half-length of the wormhole tube.

At first glimpse, one might flip both the ingoing and outgoing waveform to the other side of the branch cut while evaluating Green's function's contribution.
However, as shown in Appx.~\ref{appA}, it leads to a trivial power law irrelevant to the specific form of the effective potential.
The subtlety is that appropriately evaluating the Wronskian at a given coordinate implies a normalization of the waveform. 
It can be shown that the frequency-domain Green's function now possesses the form
\bqn
\widetilde{G}_\pm(\omega,x,y) = \frac{1}{D_{23\pm}}\frac{f_\pm(\omega,y) g_\pm(\omega,x)}{W(g_\pm,h_3)} \label{Gtilde_pm3},
\eqn
where the normalization
\bqn
D_{23\pm}=-\frac{W\left(f_\pm,h_1\right)}{W_2} ,
\eqn
and
\bqn
W_2=2i\varpi ,
\eqn
where one utilizes the subscripts ``$\pm$'' to denote the waveforms evaluated on the right (left) side of the imaginary frequency axis, $g$ and $f$ to indicate the ingoing and outgoing waveforms in regions 1 and 3, $h_{1,3}$ are the two independent waveforms in region 2 related to $g$ and $f$ while satisfying the junction conditions at the boundaries
\bqn
\left.\left[\ln g(\omega,x)\right]'\right|_{x=x_c} &=& \left.\left[\ln h_1(\omega,x)\right]'\right|_{x=x_c} ,\nb\\
\left.\left[\ln f(\omega,x)\right]'\right|_{x=-x_c} &=& \left.\left[\ln h_3(\omega,x)\right]'\right|_{x=-x_c}.\label{junCond3}
\eqn
By comparing Eq.~\eqref{Gtilde_pm3} against Eq.~\eqref{Gtilde_pm}, one has placed the source on the other side of the throat (region 3) and evaluated the Wronskian at the boundary between regions 1 and 2.
Therefore, it is meaningful also to verify whether the results remain unchanged if the initial perturbation is placed in a different region.
As discussed in Appx.~\ref{appA}, the invariance of the result is guaranteed by the normalization factor $D_{23\pm}$.

By evaluating the difference between the two sides of the branch cut, one finds
\bqn
\Delta\widetilde{G} 
= {W_2}\left[\frac{h_3f_+W\left(g_+,g_-\right)}{W\left(f_+,h_1\right)W\left(g_+,h_3\right)W\left(g_-,h_3\right)}+\frac{h_1g_-W\left(f_+,f_-\right)}{W\left(f_+,h_1\right)W\left(f_-,h_1\right)W\left(g_-,h_3\right)}\right] .\label{DeltaGTilde3}
\eqn
We relegate the somewhat tedious details of the derivations of Eqs.~\eqref{Gtilde_pm3} and~\eqref{DeltaGTilde3} to Appx.~\ref{appA}.

To discuss the specific properties of the resulting late-time tail, we plug in the specific forms of the wavefunctions and Wronskians 
\bqn
W(g_+,g_-) &=& 4i\sigma \sin\nu_1\pi ,\nb\\
W(f_+,f_-) &=& 4i\sigma \sin\nu_2\pi ,\nb\\
g_\pm(-i\sigma, x)&\propto& \sigma^{-\nu_1} , \nb\\
f_\pm(-i\sigma, x)&\propto& \sigma^{-\nu_2} , \nb\\
W(g_\pm,h_3)&=& \left[h_3'(-i\sigma,x_c)+\frac{\nu_1}{x_c}h_3(-i\sigma,x_c)\right]g_\pm(x_c) \propto \sigma^{-\nu_1} ,\nb\\
W(f_\pm,h_1)&=& \left[h_1'(-i\sigma,-x_c)-\frac{\nu_2}{x_c}h_1(-i\sigma,-x_c)\right]f_\pm(-x_c) \propto \sigma^{-\nu_2} .\label{assForms}
\eqn
Subsequently, one finds
\bqn
G(t,x,y) \propto  t^{-2\left[\mathrm{min}(\nu_1,\nu_2)+1\right]} .\label{PLTail}
\eqn

Apparently, the late-time tail is identical to that of the black hole metric when $\nu_1=\nu_2=\nu$.
However, when $\nu_1\ne \nu_2$, it is governed by the power decreasing slower, which might be ``hidden'' behind the throat.
In other words, the late-time tail phenomenon might be potentially used to distinguish the wormhole metric from its counterpart black hole.

\begin{figure}[h]
\begin{tabular}{cc}
\begin{minipage}{250pt}
\centerline{\includegraphics[width=1.0\textwidth]{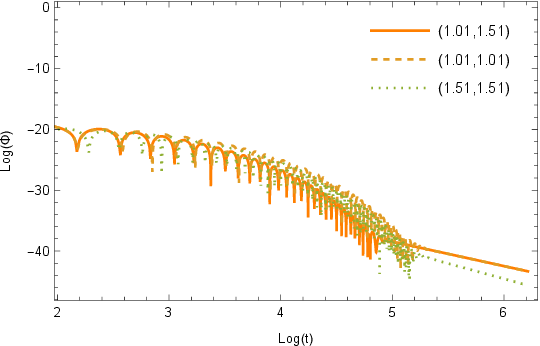}}
\end{minipage}
&
\begin{minipage}{250pt}
\centerline{\includegraphics[width=1.0\textwidth]{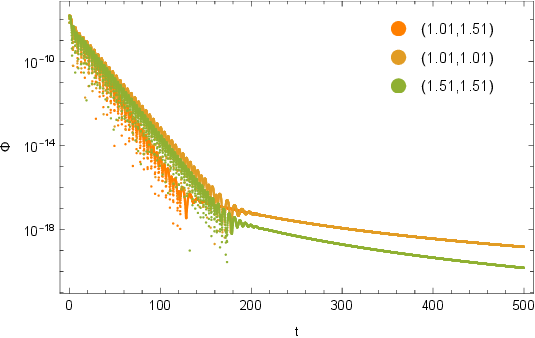}}
\end{minipage}
\end{tabular}
\renewcommand{\figurename}{Fig.}
\caption{An illustration of quasinormal oscillations and late-time tail in a Damour-Solodukhin-type wormhole in log-log (left) and log-linear (right) scales. 
The results are evaluated numerically for the effective potential given by Eq.~\eqref{Veff_DS} for different parameters $\nu_1=1.01, \nu_2=1.51$ (solid vermillion), $\nu_1=1.01, \nu_2=1.01$ (dashed orange), and $\nu_1=1.51, \nu_2=1.51$ (dotted green).
The slopes of the late-time tails shown in the log-log scales are manifestly 4.02 (vermillion and orange) and 5.02 (green).}
\label{fig_tails} 
\end{figure}

\begin{figure}[h]
\begin{tabular}{cc}
\begin{minipage}{250pt}
\centerline{\includegraphics[width=1.0\textwidth]{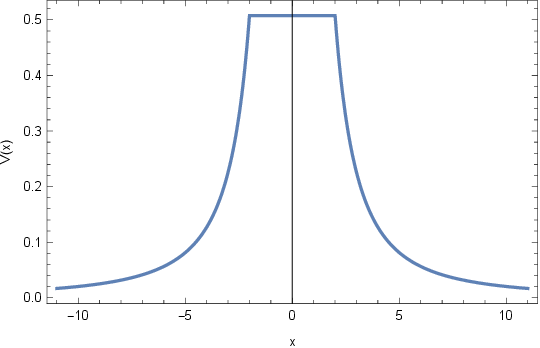}}
\end{minipage}
&
\begin{minipage}{250pt}
\centerline{\includegraphics[width=1.0\textwidth]{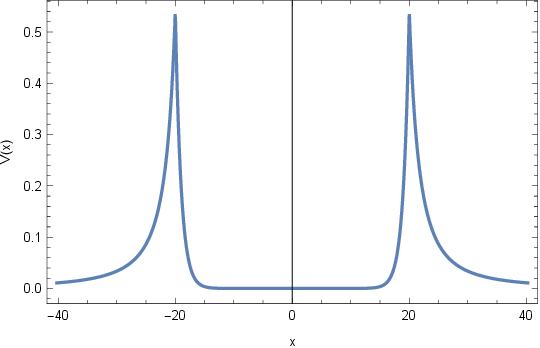}}
\end{minipage}
\end{tabular}
\renewcommand{\figurename}{Fig.}
\caption{The effective potentials utilized for numerical simulations shown in Figs.~\ref{fig_tails} and~\ref{fig_echoes}.
Left: The effective potential governed by Eq.~\eqref{Veff_DS}, the effective potential decreases in both directions according to an inverse square power form and features a plateau in the middle.
Right: The effective potential given by Eq.~\eqref{Veff_DS2}, the left and right regions of the effective potential are further separated, and an exponential suppression is introduced in the middle region.
The separation is to produce a desirable value for the resulting echo period.
While the suppression in the effective potential will not intervene with the could-be late-time tail, it effectively increases the echo signal's strength.}
\label{fig_Veffs} 
\end{figure}

These results are illustrated using the finite difference method~\cite{agr-qnm-finite-difference-01, agr-qnm-finite-difference-02}, where the resulting time evolutions for effective potentials with different parameters $(\nu_1,\nu_2)$ are shown in Fig.~\ref{fig_tails}.
In the numerical calculations, we take $\nu_1=1.01$, $\nu_2=1.51$, $x_c \simeq 2$, and $V_c \simeq 0.51$.
We show the absolute value of the waveform $\Phi = |\Psi|$ in both linear and logarithmic scales.
The initial perturbation is placed outside of the wormhole's throat.
The effective potential used in the calculations is illustrated in the left panel of Fig.~\ref{fig_Veffs}.
It is apparent that when the two power indices on the two sides of the throat are different, both power laws persist in the Green's function~\eqref{DeltaGTilde3}.
Nonetheless, only the tail index that falls the slowest is observable. 

\section{Echoes as a novel branch of asymptotic QNMs}\label{sec4}

Another essential feature in Damour-Solodukhin wormholes is the echoes.
As discussed above, a general recipe for echoes from exotic compact objects was proposed in~\cite{agr-qnm-echoes-15}.
In this section, after giving a brief account of the scheme, we explicitly evaluate the reflection amplitude in the context of a wormhole metric and assess the resulting novel branch of the QNM spectrum and the echoes. 

\subsection{Echoes in exotic compact objects}

In what follows, we briefly review Green's function approach for echoes in exotic compact objects in the context of wormhole metric primarily based on~\cite{agr-qnm-echoes-15, agr-qnm-echoes-20}.
In a Damour-Solodukhin wormhole, the event horizon of the original black hole metric is replaced by a high-tension thin shell with vanishing energy density.
The effective potential near the throat essentially has the shape of a flat valley and vanishes $V_c \sim 0$, and the throat's coordinate will be chosen as $x=0$.
At the throat, when compared to the black hole metric, the ingoing waveform will pick up a small fraction of the outgoing one, while the outgoing waveform remains unchanged.
Specifically, we have
\bqn
f_3(\omega,x) &=& f_1(\omega,x)+\widetilde{\mathcal{C}}(\omega)g_1(\omega,x) ,\label{h1Cform}
\eqn
namely, a combination of the ingoing and outgoing waveforms, whose asymptotical forms are given by Eqs.~\eqref{Asf1} and~\eqref{Asg1}.
As discussed in Appx.~\ref{appB}, as the effective potential of the original black hole is shifted, the waveforms are also measured w.r.t. the coordinate $x_c$.

It is reasonable to assume that at the vicinity of the wormhole's throat $|x| < x_c$, the effective potential is insignificant $V=V_c\sim 0$, one can approximately write down the ingoing waveform $f_3$ as a linear combination of two plane waves 
\bqn
f_3(\omega,x)\propto e^{-i\omega (x-x_c)} +\widetilde{\mathcal{R}}(\omega)e^{i\omega (x-x_c)} ,\label{h1Rform}
\eqn
for $x\to \ \mathrm{throat}$, where the reflection amplitude $\widetilde{\mathcal{R}}$ is dictated mainly by the specific nature of the compact object.

In particular, we have the relation
\bqn
\widetilde{\mathcal{C}}(\omega) = \frac{\widetilde{\mathcal{T}}_\mathrm{BH}(\omega)\widetilde{\mathcal{R}}(\omega)}{1-\widetilde{\mathcal{R}}_\mathrm{BH}(\omega)\widetilde{\mathcal{R}}(\omega)} ,\label{relCR}
\eqn
where the black hole's reflection and transmission amplitudes are defined by Eq.~\eqref{RefTransA}.
To derive Eq.~\eqref{relCR}, one makes use of the asymptotical forms given by Eqs.~\eqref{master_bc1} and~\eqref{master_bc2} at $x\to \ \mathrm{throat}$, then compares Eq.~\eqref{h1Cform} against Eq.~\eqref{h1Rform} and equates the ratios of the coefficients before the terms $e^{\pm i\omega x}$.

One plugs $h_3$ and $g_1$ into Eq.~\eqref{FormalGreen} to assess the poles of Green's function.
Apart from that, our focus is on the poles, not the branch cut of the resulting Green's function; essentially, the same procedure is performed in Eq.~\eqref{Gtilde_pm3}.
For $y\le x_c \le x$, we have
\bqn
\widetilde{G}(\omega,x,y) 
&=& \frac{f_3(\omega,y) g_1(\omega,x)}{W(g_1,f_3)}\nb\\
&=& \frac{f_1(\omega,y) g_1(\omega,x)}{W(g_1,f_3)}+\widetilde{\mathcal{C}}(\omega)\frac{g_1(\omega,y) g_1(\omega,x)}{W(g_1,f_3)}\nb\\
&=& \frac{f_1(\omega,y) g_1(\omega,x)}{W(g_1,f_1)}+\widetilde{\mathcal{C}}(\omega)\frac{g_1(\omega,y) g_1(\omega,x)}{W(g_1,f_1)}\nb\\
&\equiv& \widetilde{G}_\mathrm{BH}(\omega,x-x_c,y-x_c) +\widetilde{\mathcal{C}}(\omega)\frac{g_1(\omega,y) g_1(\omega,x)}{W_\mathrm{BH}} ,\label{Gtilde_h3}
\eqn
where $W_\mathrm{BH}=W(g_\mathrm{BH},f_\mathrm{BH})=W(g_1,f_1)$ and $\widetilde{G}_\mathrm{BH}$ are the Wronskian and Green's function of the original black hole metric (c.f. Eqs.~\eqref{master_bc1} and~\eqref{master_bc2}).
In other words, the QNMs of the original black hole mainly persist in the wormhole metric, and echoes are attributed to the poles in $\widetilde{\mathcal{C}}(\omega)$ defined by Eq.~\eqref{relCR}.

From a rather generic perspective, the presence of echoes can be attributed to the phase shift of the reflection amplitude~\cite{agr-qnm-echoes-15, agr-qnm-echoes-20, agr-strong-lensing-correlator-15}.
Before delving into specific derivations, it is essential to note that the definitions of $\widetilde{\mathcal{C}}$ and $\widetilde{\mathcal{R}}$ depend explicitly on the choice of $x_c$.
In this paper, we have made a choice so that Eq.~\eqref{relCR} does not have an explicit dependence on $x_c$.
This is intended so that the occurrence of echoes can {\it not} be attributed to some specific choice of coordinates (c.f. Eqs.~\eqref{altDefR} and~\eqref{relCR0} below in Appx.~\ref{appB} and the discussions therein).

On the one hand, as the original black hole metric does not imply any echo phenomenon, it is reasonable first to assume that the quantities $\widetilde{\mathcal{T}}_\mathrm{BH}$ and $\widetilde{\mathcal{R}}_\mathrm{BH}$ are moderate functions of the frequency $\omega$ and can be treated as constants.
On the other hand, if the reflection amplitude possesses the form
\bqn
\widetilde{\mathcal{R}}(\omega)=\widetilde{\mathcal{R}}_0 e^{i\omega L} ,\label{RForm2echo}
\eqn
where $L$ is a constant related to $x_c$, it is straightforward to show that Eq.~\eqref{relCR} can largely be viewed as the inverse Fourier transform of
\bqn
\mathcal{C}(t) = \int_{-\infty}^{+\infty} \frac{d\omega}{2\pi}\widetilde{\mathcal{C}}(\omega)e^{-i\omega t}
=A\left[\delta(t)+B\delta(t+T)+B^2\delta(t+2T)\cdots\right],\label{InvrelCR}
\eqn
where
\bqn
A &=& -\frac{\widetilde{\mathcal{T}}_\mathrm{BH}}{\widetilde{\mathcal{R}}_\mathrm{BH}},\nb\\
B &=& \frac{1}{\widetilde{\mathcal{R}}_\mathrm{BH}\widetilde{\mathcal{R}}_0},\nb\\
T &=& L.
\eqn
Subsequently, the time domain profile governed by the second term on the r.h.s. of Eq.~\eqref{Gtilde_h3} is the quasinormal oscillation governed by $W_\mathrm{BH}$ on the demonstrator modulated by the echoes given by Eq.~\eqref{InvrelCR}. 
Specifically, we have the following convolution that leads to echoes
\bqn
G(t,y,x) \sim \int d\tau \mathcal{H}(t) \mathcal{C}(t-\tau)
=\mathcal{H}(t)+\mathcal{H}(t+T)+\mathcal{H}(t+2T)+\cdots,
\eqn
where
\bqn
\mathcal{H}(t) = \int_{-\infty}^{+\infty} \frac{d\omega}{2\pi}\frac{g_1(\omega,y) g_1(\omega,x)}{W_\mathrm{BH}}
\eqn
is characterized by the quasinormal oscillations of the original black hole.

It is worth noting that, under the above circumstances, the poles of Eq.~\eqref{relCR} are evenly distributed and lying parallel to the real frequency axis~\cite{agr-qnm-echoes-15, agr-qnm-echoes-16, agr-qnm-echoes-20}.
This is because if $\omega$ is a pole of Eq.~\eqref{relCR}, it implies that $\omega+2n\pi/T$ (where $n$ is an arbitrary integer) is also a pole.
The union of such poles corresponds to an additional branch of the QNM spectrum of Green's function~\eqref{Gtilde_h3}.
We note that such a feature also occurs due to the spectral instability~\cite{agr-qnm-instability-07}, which might be triggered by ultraviolet perturbations such as discontinuity~\cite{agr-qnm-lq-03}.
The gap between this branch of the spectrum and the real axis governs whether the echoes are observationally feasible.
The distance between successive poles defines the echo period.

\subsection{Reflection amplitude at the wormhole's throat}

It is interesting to note that for the case of Damour-Solodukhin wormholes, the reflection amplitude can be evaluated explicitly.
The following derivation uses the transfer matrix, which closely follows the arguments given in~\cite{agr-qnm-echoes-16}.

When viewed as a scattering problem, the effective potential of the original black hole metric $V_\mathrm{eff}^\mathrm{BH}(x)$ corresponds to a $2\times 2$ transfer matrix $T$.
Using Eqs.~\eqref{master_bc1} and~\eqref{fluxConv}, it is found to be
\bqn
T 
=\begin{pmatrix}A_\mathrm{in}^*&A_\mathrm{out}\\A_\mathrm{out}^*&A_\mathrm{in}\end{pmatrix}
= \begin{pmatrix}B_\mathrm{out}^*&-B_\mathrm{in}^*\\-B_\mathrm{in}&B_\mathrm{out}\end{pmatrix} , \label{TBHmatrix}
\eqn
which satisfies
\bqn
\begin{pmatrix}A_\mathrm{R}'\\A_\mathrm{L}'\end{pmatrix}
= T \begin{pmatrix}A_\mathrm{R}\\A_\mathrm{L}\end{pmatrix},
\eqn
where
\begin{equation}
\Psi_\mathrm{RW}(x) = 
\begin{cases}
    A_\mathrm{R}'e^{i\omega x} + A_\mathrm{L}'e^{-i\omega x} &\ \ \ \mathrm{for}\ \ \ \  x\to +\infty,\\
    A_\mathrm{R}e^{i\omega x} + A_\mathrm{L}e^{-i\omega x} &\ \ \ \mathrm{for}\ \ \ \  x\to -\infty . 
\end{cases}  
\end{equation}

It can be shown that the corresponding transfer matrix for an observable located at $x=0$ to receive an incident wave from $x\to-\infty$ is
\bqn
\mathbb{T}^\mathrm{lhs}=\begin{pmatrix}e^{ i \omega x_c}&0\\0&e^{- i \omega x_c}\end{pmatrix} \sigma T^{-1}\sigma ,
\eqn
where
\bqn
\sigma = \begin{pmatrix}0&1\\1&0\end{pmatrix} .
\eqn
In particular, the transfer matrix for the spatially reflected metric $V_\mathrm{eff}^\mathrm{BH}(-x)$ is $\sigma T^{-1}\sigma$.
This is because the replacement $x\to -x$ in Eq.~\eqref{master_bc2} implies exchanging the incident and scattered wavefunctions ($T^{-1}$) as well as switching the column vector's two components ($\sigma$) twice.
Besides, the effective potential $V_\mathrm{eff}^\mathrm{BH}(-x)$ must be displaced to the left by $x_c$, which is implemented by the phase shift $e^{ i \omega x_c}$ in the diagonal matrix.

According to the definition~\eqref{h1Rform}, the wormhole's reflection amplitude is found to be
\bqn
\widetilde{\mathcal{R}}(\omega)
=\frac{\mathbb{T}^\mathrm{lhs}_{12}e^{-i\omega(x-x_c)}}{\mathbb{T}^\mathrm{lhs}_{22}e^{+i\omega(x-x_c)}} =\widetilde{\mathcal{R}}_\mathrm{BH}e^{4i\omega x_c} .\label{TildeRForm}
\eqn
Unlike Eq.~\eqref{RefTransA}, this reflection amplitude corresponds to the scattering of an ingoing wave expressed w.r.t. to the coordinate $x=x_c$.
In the above derivation, we have used Eqs.~\eqref{TBHmatrix},~\eqref{RefTransA}, and~\eqref{fluxConv}.
When compared against Eq.~\eqref{RForm2echo}, we find the echo period 
\bqn
T = L=4x_c .\label{echoPeriodWH}
\eqn
It measures the period for a wavepacket to bounce between $-x_c$ and $x_c$.

By substituting Eq.~\eqref{TildeRForm} into Eq.~\eqref{relCR}, one finds
\bqn
\widetilde {\mathcal C}(\omega)
=\frac{\widetilde{\mathcal{T}}_\mathrm{BH}\widetilde{\mathcal{R}}_\mathrm{BH}}{e^{-4i \omega x_c} - \widetilde{\mathcal{R}}_\mathrm{BH}^2} . \label{TildeCForm}
\eqn
As a further confirmation, an alternative approach to derive Eqs.~\eqref{TildeRForm} and~\eqref{TildeCForm} is given in Appx.~\ref{appB}.

\begin{figure}[h]
\begin{tabular}{cc}
\begin{minipage}{250pt}
\centerline{\includegraphics[width=1.0\textwidth]{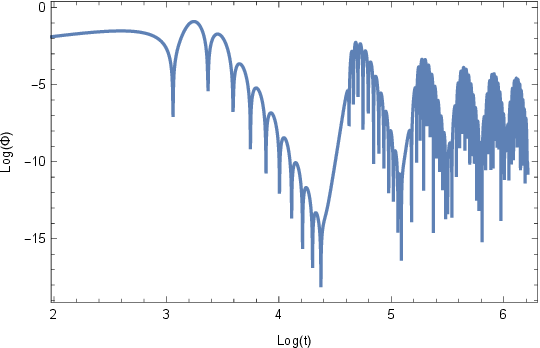}}
\end{minipage}
&
\begin{minipage}{250pt}
\centerline{\includegraphics[width=1.0\textwidth]{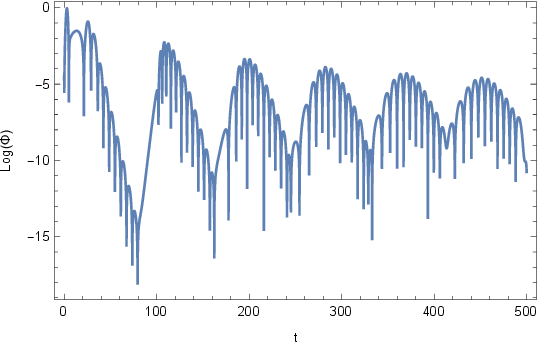}}
\end{minipage}
\end{tabular}
\renewcommand{\figurename}{Fig.}
\caption{An illustration of echoes in a Damour-Solodukhin-type wormhole in log-log (left) and semilog (right) scales. 
The echo period shown in the semilog scale is manifestly 80.}
\label{fig_echoes} 
\end{figure}

The numerical simulations are presented in Fig.~\ref{fig_echoes} using the following modified form
\begin{equation}
V^\mathrm{DS}_\mathrm{eff}(x) = 
\begin{cases}
   V^\mathrm{BH}_\mathrm{eff}(x-x_0, \nu)   &\ \ \mathrm{region}\ 1 \ \ \ x > x_c, \\
   {V_c}\cosh(x)   &\ \ \mathrm{region}\ 2 \ \ \ -x_c\le x \le x_c, \\
   V^\mathrm{BH}_\mathrm{eff}(-(x-x_0),\nu)   &\ \ \mathrm{region}\ 3 \ \ \ x < -x_c, 
\end{cases}  \label{Veff_DS2}
\end{equation}
for the effective potential, where $\nu = 1.01$, $x_c=20$, $V_c=0.54\ \mathrm{sech}(x_c)\simeq 2.23\times 10^{-9}$, and $x_0 = 50/3$.
The shape of the effective potential is shown in the right panel of Fig.~\ref{fig_Veffs}.
For numerical convenience, the effective potential in region 2 is suppressed exponentially by a hyperbolic $\cosh(x)$ function, and $V^\mathrm{BH}_\mathrm{eff}(x)$ is shifted by $x_0$ in both directions.
Compared to Eq.~\eqref{Veff_DS}, the modification to the effective potential does not affect the physical essence but is due to numerical convenience.
The separation is introduced to produce a value for the echo period $T=4x_c=80$, chosen to be more significant than that of the typical QNMs so that they are modulated by the echo arches.
While the exponential suppression will not intervene with any feature of the could-be late-time tail, it effectively increases the echo signal’s strength.

\section{An interplay between echoes and late-time tail}\label{sec5}

\begin{figure}[h]
\begin{tabular}{cc}
\begin{minipage}{250pt}
\centerline{\includegraphics[width=1.0\textwidth]{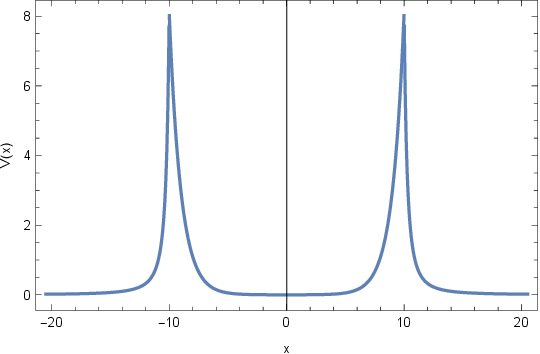}}
\end{minipage}
&
\begin{minipage}{250pt}
\centerline{\includegraphics[width=1.0\textwidth]{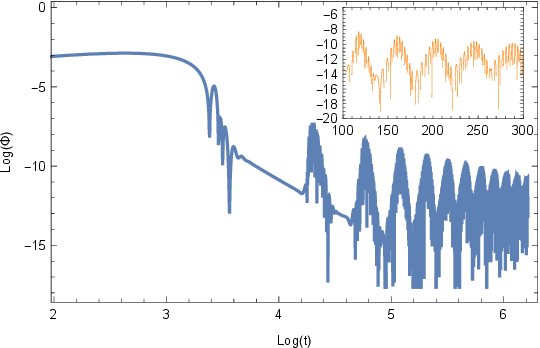}}
\end{minipage}
\end{tabular}
\renewcommand{\figurename}{Fig.}
\caption{An illustration of echoes sitting on top of the late-time tail in a Domour-Solodukhin-type wormhole.
Left: The effective potential utilized for the numerical simulations.
Right: The obtained time evolution of the waveform is shown in log-log and semilog (inset) scales.
The echo period shown in the semilog scale is 40, while the slope of the late-time tails shown in the log-log scale is manifestly 4.02.}
\label{fig_echoes_on_tail} 
\end{figure}

As discussed, both echoes and tails are late-time phenomena, and in the case of a Damour-Solodukhin-type wormhole, they might appear simultaneously.
The results shown in Figs.~\ref{fig_tails} and~\ref{fig_echoes} only illustrate one phenomenon at a time, even though both underlying mechanisms seem relevant.
For instance, as echoes are essentially a collective effect of minor quasinormal oscillations, they might have already been suppressed before the late-time tail becomes visible.
We note that such an interplay between the two phenomena might not be straightforward.
Nonetheless, we show in Fig.~\ref{fig_echoes_on_tail} the intertwined late-time tails and echoes by adequately tuning the metric parameters.

The numerical simulations presented in Fig.~\ref{fig_echoes} use the effective potential given by Eq.~\eqref{Veff_DS2} with the parameters $\nu = 1.01$, $x_c=10$, $V_c=0.00074$, and $x_0 = 9.5$.
The shape of the effective potential is shown in the left panel of Fig.~\ref{fig_echoes_on_tail}, while the time evolution is presented in the right panel.
From the inset, it is apparent that the echo arches are constituted by quasinormal oscillations with smaller periods.
The echo period is found to be $T= 4x_c=40$.
The late-time tail is also manifestly shown in the log-log scale by an inverse power law whose slope is 4.02.

It is observed numerically that as the potential wall's height gradually decreases, the emergence of the late-time tail is delayed and ultimately ceases to appear. 
This result may initially seem counterintuitive, as the QNMs, which correspond to transient states confined within the potential well, should gradually transform into normal modes when the wall's height approaches infinity. 
Consequently, it would take an infinite amount of time for the quasinormal oscillations to diminish, rendering the late-time tail unobservable. 
However, further analysis reveals two factors at play: as the wall's height increases, its thickness decreases. 
For a specific height, a reduction in thickness destabilizes the QNMs and amplifies the magnitude of the frequency's imaginary part. 
In our calculations, we have adjusted the model's parameters so that the time scale for the fundamental mode to be overtaken by the tail aligns with the emergence of the echoes.

\begin{figure}[h]
\begin{tabular}{cc}
\begin{minipage}{250pt}
\centerline{\includegraphics[width=1.0\textwidth]{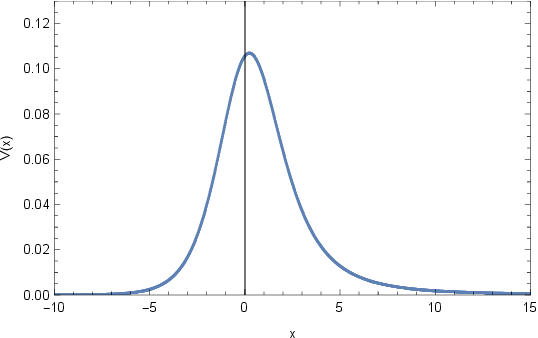}}
\end{minipage}
&
\begin{minipage}{250pt}
\centerline{\includegraphics[width=1.0\textwidth]{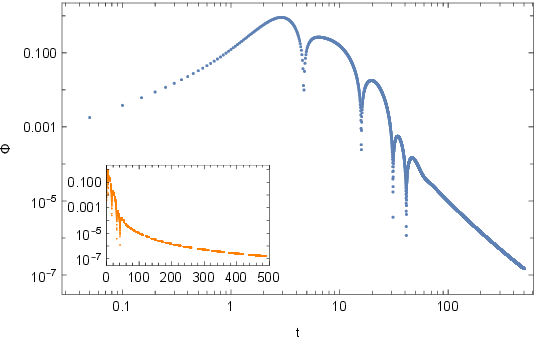}}
\end{minipage}
\\
\begin{minipage}{250pt}
\centerline{\includegraphics[width=1.0\textwidth]{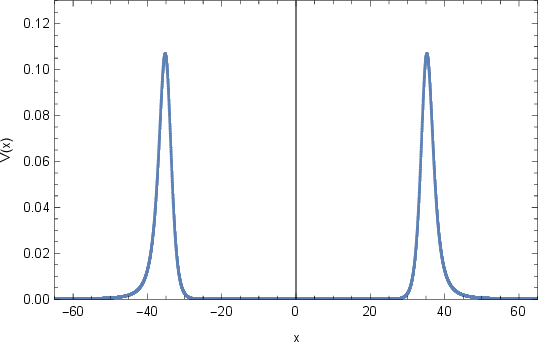}}
\end{minipage}
&
\begin{minipage}{250pt}
\centerline{\includegraphics[width=1.0\textwidth]{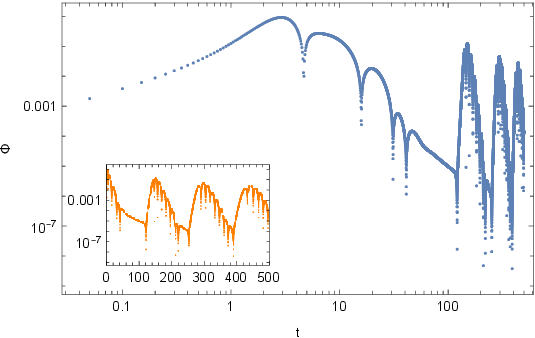}}
\end{minipage}
\end{tabular}
\renewcommand{\figurename}{Fig.}
\caption{Late-time tail and echoes sitting on top of the tail in a Schwarzschild Damour-Solodukhin wormhole.
Left column: The effective potentials utilized for the numerical simulations.
Right column: The obtained time evolutions of the waveforms shown in log-log and semilog (inset) scales.
The echo period shown in the semilog scale is 140, while the slope of the late-time tails shown in the log-log scale is 2.02.}
\label{fig_echoes_on_tail_RW} 
\end{figure}

Before closing this section, we argue that the approximated potentials Eqs.~\eqref{Veff_alpha} and~\eqref{Veff_DS} investigated in this work are pertinent to addressing specific Damour-Solodukhin wormholes. 
Because one does not possess closed analytic forms for the ingoing and outgoing waveforms, similar approximations have been adopted~\cite{agr-qnm-tail-05, agr-qnm-tail-06, agr-qnm-tail-12, agr-qnm-tail-13, agr-qnm-tail-20, agr-qnm-tail-21, agr-qnm-tail-22, agr-qnm-tail-31, agr-qnm-tail-32, agr-qnm-tail-33, agr-qnm-tail-39} to capture the dominant behavior of the effective potential at spatial infinity, which is relevant to the outgoing wave. 
For the ingoing wave for a black hole metric, as it does not contain any branch cut, as long as the effective potential decreases faster than an exponential function.
Subsequently, the specific forms of the effective potential and the ingoing waveform do not affect the shape of the power-law tail~\cite{agr-qnm-tail-39}. 
In the case of black holes, the results based on such approximation carried out for specific black hole metrics, such as the Regge-Wheeler potential, have shown to agree well with the numerical results. 
To demonstrate the adequacy of our approach for wormholes, we illustrate a realistic metric by numerically showcasing the coexistence of the late-time tail and echoes in the Schwarzschild Damour-Solodukhin wormhole. 
The results are given in Fig.~\ref{fig_echoes_on_tail_RW}. 
The calculations are performed by considering initial scalar perturbations near a wormhole constructed by the Regge-Wheeler potential Eq.~\eqref{f_RW} with the following parameters
\bqn
\ell &=& 0.01 ,\nb\\
s &=& 0,\nb\\
r_h &=& 1.
\eqn
The effective potential presented in the tortoise coordinate $x$ is shown in the top-left panel of Fig.~\ref{fig_echoes_on_tail_RW}.
The quasinormal oscillations and late-time tail are shown in the top-right panel, where the slope of the tail is found to be $2(\ell+1)=2.02$, in agreement with the analytic finding in Eq.~\eqref{PLTail}.
The effective potential is then mirrored and separated by a distance $2x_c = 70$, as indicated by the bottom-left panel.
As shown in the bottom-right panel, the resulting temporal profile constitutes echoes sitting on top of the late-time tail.
It is observed that the echoes arrive at a time scale comparable with the emergence of the late-time tail.
The choice of parameters guarantees that the echo period is more significant than that of the quasinormal oscillations.
We note that before the onset of echoes, the two plots shown in the right column of Fig.~\ref{fig_echoes_on_tail_RW} are essentially identical, which is consistent with causality arguments, based on which the emergence of echoes is interpreted as the initial perturbations traversing through the wormhole's throat and bouncing back.
The results shown in Fig.~\ref{fig_echoes_on_tail_RW} are qualitatively in agreement with Fig.~\ref{fig_echoes_on_tail} discussed above.

\section{Concluding remarks}\label{sec6}

In conclusion, this work explored the concept of Damour-Solodukhin wormholes, which intriguingly share several characteristics with black holes. 
Our study has been specifically focused on two unique aspects of the waveforms emitted from these wormholes: the late-time tails and the echoes.
It is argued that these elements are crucial in differentiating wormholes from black holes.
Though often elaborated separately in the literature, these two phenomena both emerge in the later phases of quasinormal oscillations and originate from the singularities in the Green’s function.
The late-time tail phenomenon arises due to branch cuts in the Green’s function.
For the wormhole method, such a feature implies a modification of the existing mathematical recipe used for black hole metrics. 
In particular, within the framework of Damour-Solodukhin wormholes, the singularities persist in both the ingoing and outgoing waveforms. 
In contrast, the echoes are identified as resulting from a novel set of quasinormal poles, which are additional to those found in corresponding black holes. 
This is reminiscent of cases where the spacetime metric exhibits a discontinuity.

Our analysis indicates that both the late-time tails and echoes are observationally significant in distinguishing a wormhole from a black hole. 
The inverse power law of the late-time tail, as given by Eq.~\eqref{PLTail}, suggests the possibility of inferring spacetime properties hidden behind the wormhole throat. 
Furthermore, we have proposed a scenario where both phenomena coexist.
In order to assess whether the interplay of echoes and late-time tails is relevant to future space-borne gravitational wave detectors, further studies are warranted to estimate the signal-to-noise ratio using realistic detector parameters. 
This could potentially pave the way for new insights into the nature of these enigmatic astrophysical objects.

\section*{Acknowledgements}

WLQ is grateful for the hospitality of Huazhong University of Science and Technology, where part of the manuscript was prepared.
We gratefully acknowledge the financial support from Brazilian agencies 
Funda\c{c}\~ao de Amparo \`a Pesquisa do Estado de S\~ao Paulo (FAPESP), 
Funda\c{c}\~ao de Amparo \`a Pesquisa do Estado do Rio de Janeiro (FAPERJ), 
Conselho Nacional de Desenvolvimento Cient\'{\i}fico e Tecnol\'ogico (CNPq), 
and Coordena\c{c}\~ao de Aperfei\c{c}oamento de Pessoal de N\'ivel Superior (CAPES).
This work is supported by the National Natural Science Foundation of China (NSFC).
A part of this work was developed under the project Institutos Nacionais de Ci\^{e}ncias e Tecnologia - F\'isica Nuclear e Aplica\c{c}\~{o}es (INCT/FNA) Proc. No. 464898/2014-5.
This research is also supported by the Center for Scientific Computing (NCC/GridUNESP) of S\~ao Paulo State University (UNESP).

\appendix

\section{Normalization of the waveform}\label{appA}

In this appendix, we give an account of the derivation of Eqs.~\eqref{Gtilde_pm3} and~\eqref{DeltaGTilde3} given in the main text.

First, we point out that one cannot flip both the ingoing and outgoing waveform simultaneously to the branch cut's other side in order to assess the contribution of Green's function, specifically
\bqn
\widetilde{G}_\pm(\omega,x,y) \stackrel{?}{=} \frac{f_\pm(\omega,y) g_\pm(\omega,x)}{W(f_\pm, g_\pm)}. \nb
\eqn
To evaluate the difference across the branch cuts, one uses the asymptotical forms in Eq.~\eqref{assForms} and notices
\bqn
W(f_\pm ,g_\pm) = \sigma^{-(\nu_1+\nu_2)} ,\nb
\eqn
instead of a form similar to the first two rows of Eq.~\eqref{assForms}, which can be verified by using the explicit forms Eqs.~\eqref{gH2} and~\eqref{fH3}.
This gives the asymptotical contribution
\bqn
G(t)\propto \int d\sigma \Delta \widetilde{G}(\omega) e^{-i\omega t} \stackrel{?}{\sim} t^{-1} ,
\eqn
which is irrelevant to the specific form of the effective potential. 

To understand why the above result is incorrect (besides its apparent contradiction to the numerical calculations), we proceed with a discussion regarding the normalization of the waveform in an attempt to reproduce the known result given by Eq.~\eqref{Gtilde_pm} from a different perspective.
In deriving Eq.~\eqref{Gtilde_pm}, one assumes that the source of the Green's function is located at $y \le x_c \le x$.
The Wronskian on the denominator is a quantity independent of the coordinate.
However, if one tries to evaluate it in the region $x \ge x_c$, a {\it paradox} appears.
For $x \ge x_c$, the outgoing waveform $g_1(\omega, x)$ remains to be that given by Eq.~\eqref{gH2}, where we use the subscript to denote the spatial region.
On the other hand, the ingoing waveform is a linear combination of two solutions of the homogeneous equation, namely, $g_1$ and 
\bqn
f_1(\omega, x)=e^{-i\left(\rho+\frac12\right)\frac{\pi}{2}}\sqrt{\frac{\pi\omega x}{2}}H_\rho^{(2)}(\omega x). 
\eqn
We denote 
\bqn
f_2(\omega,x)=f_1(\omega,x)+\kappa_1 g_1(\omega,x) ,\nb
\eqn
since $f_1$ is asymptotically an ingoing wave and $\kappa_1$ mixes into the waveform by a small fraction of $g_1$.

Subsequently, the Wronskian reads
\bqn
W(g_2,f_1)=W(g_1,f_1)
=\left(\sqrt{\frac{\pi\omega x}{2}}\right)^2 W\left(H_\rho^{(1)},H_\rho^{(2)}\right)
=2i\omega ,\nb
\eqn
which is different from the third line of Eq.~\eqref{assForms}.
To resolve the apparent contradiction, we note that the expression of $f_2(\omega, x)$ is not properly normalized. It should be
\bqn
f_2(\omega,x)=\frac{1}{C_{12}}\left[f_1(\omega,x)+\kappa_1 g_1(\omega,x)\right] ,
\eqn
where both the normalization factor $C_{12}$ and combination coefficient $\kappa_1$ are determined by the connection condition at $x=x_c$.

To be more specific, let us explicitly consider a tractable scenario that $V=V_c=V_\mathrm{eff}^\mathrm{BH}(x_c)$ for region 2: $x\le x_c$ in Eq.~\eqref{Veff_alpha}.
Then, it is straightforward to show that
\bqn
\kappa_1 &=& -\left.\frac{(f_1)'+i\varpi f_1}{(g_1)'+i\varpi g_1}\right|_{x=x_c} ,\nb\\
C_{12} &=& \left.\frac{2i\omega}{\left[(g_1)'+i\varpi g_1\right]e^{-i\varpi}}\right|_{x=x_c} , \label{kappaC}
\eqn
where $\varpi=\sqrt{\omega^2-V_c}$.
At $\omega\to 0$, where the inverse Fourier transform~\eqref{IntBranchCut} picks up the dominant contribution, the normalization plays an increasingly important role as it becomes divergent.
By plugging Eq.~\eqref{kappaC} into the Wronkian, one finds 
\bqn
W(g_2,f_1)=\frac{1}{C_{12}}W(g_1,f_1)
=\left.\left[(g_1)'+i\varpi g_1\right]\right|_{x=x_c}e^{-i\varpi} ,\label{normWron}
\eqn
which readily agrees with Eq.~\eqref{assForms}.
We note that if one does not specify the effective potential in region 2 and denote the two independent waveforms as $f_2$ and $g_2$, it is not difficult to show that $C_{12}$ has a more general form
\bqn
C_{12} &=& \frac{W_1}{W\{f_2, g_1\}} , \label{kappaCGen}
\eqn
where $W_1=W(g_1,f_1)$ and the Wronkian on the denominator must be evaluated at the junction point $x=x_c$.

The above result indicates that one can evaluate the Wronskian at any point, as expected, by adequately considering the waveform normalization.
Now, we proceed to evaluate Green's function for the effective potential defined in Eq.~\eqref{Veff_DS}, which possesses the form
\bqn
\widetilde{G}_\pm(\omega,x,y) = \frac{1}{D_{23\pm}}\frac{f_\pm(\omega,y) g_\pm(\omega,x)}{W(g_\pm,h_3)},\nb
\eqn
where one utilizes the subscripts ``$\pm$'' to denote the waveforms evaluated on the right (left) side of the imaginary frequency axis, $g$ and $f$ to indicate the ingoing and outgoing waveforms in regions 1 and 3, $h_{1,3}$ are the two independent waveforms in region 2 related to $g$ and $f$ satisfying the junction conditions \eqref{junCond3} at the boundaries.
Moreover, the ingoing wave, for which we also place the branch cut on the negative imaginary axis by adopting Eq.~\eqref{anaCon1} for its analytic extension, reads
\bqn
f(\omega, x)=e^{i(\rho+1)\frac{\pi}{2}}\sqrt{\frac{-\pi\omega x}{2}}H_\rho^{(1)}(-\omega x) , \label{fH3} 
\eqn
for $x<-x_c<0$, and the normalization is given by
\bqn
D_{23\pm}=\frac{W\left(f_\pm,h_1\right)}{W_2} ,\nb
\eqn
is the corresponding normalization factor between regions 2 and 3, reminiscent of Eq.~\eqref{kappaCGen}.
The subscripts ``$\pm$'' indicate that the waveforms are evaluated on the branch cut's right (left) side.
We consider the source of the perturbation to be located in region 3, with the corresponding ingoing wave $f_3\equiv f$, the observer in region 1, and the outgoing wave denoted by $g_1\equiv g$.
In region 2, $h_{1,3}$ are the two independent waveforms satisfying the junction conditions~\eqref{junCond3} at the two boundaries.

By comparing Eq.~\eqref{Gtilde_pm3} against Eq.~\eqref{Gtilde_pm}, one has placed the source on the other side of the throat (region 3) and evaluated the Wronskian at the boundary between regions 1 and 2.
Therefore, it is meaningful to verify whether the results remain unchanged if the initial perturbation is placed in a different region.
As discussed in Appx.~\ref{appA}, the invariance of the result is guaranteed by the normalization factor $D_{23\pm}$.

By evaluating the difference between the two sides of the branch cut, one finds the desired result
\bqn
\Delta\widetilde{G} 
&=& {W_2}\left[g_+f_+\frac{1}{W\left(f_+,h_1\right)W\left(g_+,h_3\right)}-g_-f_-\frac{1}{W\left(f_-,h_1\right)W\left(g_-,h_3\right)}\right] \nb\\
&=& {W_2}\frac{f_+W\left(f_-,h_1\right)g_+W\left(g_-,h_3\right)-f_-W\left(f_+,h_1\right)g_-W\left(g_+,h_3\right)}{W\left(f_+,h_1\right)W\left(g_+,h_3\right)W\left(f_-,h_1\right)W\left(g_-,h_3\right)} \nb\\
&=& {W_2}\frac{h_3f_+W\left(g_+,g_-\right)W\left(f_-,h_1\right)+h_1g_-W\left(f_+,f_-\right)W\left(g_+,h_3\right)}{W\left(f_+,h_1\right)W\left(g_+,h_3\right)W\left(f_-,h_1\right)W\left(g_-,h_3\right)}\nb\\
&=& {W_2}\left[\frac{h_3f_+W\left(g_+,g_-\right)}{W\left(f_+,h_1\right)W\left(g_+,h_3\right)W\left(g_-,h_3\right)}+\frac{h_1g_-W\left(f_+,f_-\right)}{W\left(f_+,h_1\right)W\left(f_-,h_1\right)W\left(g_-,h_3\right)}\right] ,\nb
\eqn
where 
\bqn
W_2=W\left(e^{2i\varpi},e^{-2i\varpi}\right)=2i\varpi .\nb
\eqn
Moreover, the Wronskians between waveforms from adjacent regions must be evaluated at the corresponding boundary.

\section{An alternative derivations of the reflection amplitude at the wormhole's throat}\label{appB}

Here, we provide a second derivation of Eqs.~\eqref{TildeRForm} and~\eqref{TildeCForm}, which is more intuitive but somewhat tedious.

To accommodate the wormhole metric, we first shift the effective potential to the right by a distance $x_c$ so that $V_\mathrm{eff}(x)\to V_\mathrm{eff}(x-x_c)$ and the wormhole's throat is placed at $x=0$.
Specifically, we change the argument $x\to x-x_c$ in the wavefunctions, and the asymptotic forms \eqref{master_bc1} and~\eqref{master_bc2} become
\bqn
f_1(\omega, x)=\left\{\begin{matrix}e^{-i \omega(x-x_c)}&x\to \ \mathrm{throat} \\A_{\mathrm{out}}e^{i\omega (x-x_c)}+A_{\mathrm{in}}e^{-i\omega (x-x_c)} &x\to +\infty\end{matrix}\right. ,\label{Asf1}
\eqn
and
\bqn
g_1(\omega, x)=\left\{\begin{matrix}e^{i\omega (x-x_c)}&x\to +\infty \\B_{\mathrm{out}}e^{i \omega (x-x_c)}+B_{\mathrm{in}}e^{-i \omega (x-x_c)} &x\to  \ \mathrm{throat}\end{matrix}\right. ,\label{Asg1}
\eqn
where we have denoted the limit for the ingoing wave at $x\to \ \mathrm{throat}$ instead of $x\to -\infty$.

Alternatively, to assess the waveform on the other end of the throat, we perform a spatial reflection $x\to -x$ to the effective potential, then apply a shift to the left $x\to x+x_c$.
We have
\bqn
g_3(\omega, x)=\left\{\begin{matrix}e^{+i \omega(x+x_c)}&x\to  \ \mathrm{throat} \\A_{\mathrm{out}}e^{-i\omega (x+x_c)}+A_{\mathrm{in}}e^{i\omega (x+x_c)} &x\to -\infty\end{matrix}\right. ,\label{Asg3}
\eqn
and
\bqn
f_3(\omega, x)=\left\{\begin{matrix}e^{-i\omega (x+x_c)}&x\to -\infty \\B_{\mathrm{out}}e^{-i \omega (x+x_c)}+B_{\mathrm{in}}e^{i \omega (x+x_c)} &x\to  \ \mathrm{throat}\end{matrix}\right. ,\label{Asf3}
\eqn
where we have mapped $f_1\to g_3$ and $g_1\to f_3$ and, again, denote the limit for the outgoing wave at $x\to  \ \mathrm{throat}$.

Now, on the one hand, we denote the ingoing wave at the wormhole's throat according to Eq.~\eqref{h1Cform} as
\bqn
f_3(\omega,x) &=& f_1(\omega,x)+\widetilde{\mathcal{C}}(\omega)g_1(\omega,x) .\nb
\eqn
By using the asymptotic form of Eqs.~\eqref{Asf1},~\eqref{Asg1}, and~\eqref{Asf3}, we have
\bqn
\frac{e^{i \omega x_c}+\tilde{\mathcal C}  B_{\mathrm{in}}e^{i \omega x_c}}{B_{\mathrm{out}} e^{-i \omega x_c}}=\frac{\tilde{\mathcal C} B_{\mathrm{out}} e^{-i \omega x_c}}{B_{\mathrm{in}} e^{i \omega x_c}} ,
\eqn
which gives
\bqn
\widetilde {\mathcal C}
=\frac{B_{\mathrm{in}}e^{2i\omega x_c}}{B^2_{\mathrm{out}}e^{-2i \omega x_c} -B^2_{\mathrm{in}}e^{2i\omega x_c}}
=\frac{\widetilde{\mathcal{T}}_\mathrm{BH}\widetilde{\mathcal{R}}_\mathrm{BH}}{e^{-4i \omega x_c} - \widetilde{\mathcal{R}}_\mathrm{BH}^2} ,\nb
\eqn
which is precisely Eq.~\eqref{TildeCForm}.

On the other hand, $f_3$ can be expressed in terms of the definition of the reflection amplitude given by Eq.~\eqref{h1Rform}
\bqn
f_3(\omega,x)\propto e^{-i\omega (x-x_c)} +\widetilde{\mathcal{R}}(\omega)e^{i\omega (x-x_c)} ,\nb
\eqn
where we note that $\widetilde{\mathcal{R}}$ is defined as the wormhole's throat $x\to 0$, but the waveforms are expressed w.r.t. $x=x_c$.
If they were written down w.r.t. the position $x=0$, we would have
\bqn
f_3(\omega,x)\propto e^{-i\omega x} +\widetilde{\mathcal{R}'}(\omega)e^{i\omega x} ,\label{altDefR}
\eqn
where
\bqn
\widetilde{\mathcal{R}'}=\widetilde{\mathcal{R}}e^{-2i\omega x_c}.
\eqn
As a result, instead of Eq.~\eqref{relCR}, we would have
\bqn
\widetilde{\mathcal{C}} = \frac{\widetilde{\mathcal{T}}_\mathrm{BH}\widetilde{\mathcal{R}'}}{e^{-2i\omega x_c}-\widetilde{\mathcal{R}}_\mathrm{BH}\widetilde{\mathcal{R}'}} ,\label{relCR0}
\eqn
which inherits the factor $e^{-2i\omega x_c}$ that gives rise to apparent echoes (with an incorrect period) primarily owing to a specific choice of the expansion point.
Also, similar results are obtained if one keeps Eq.~\eqref{h1Rform} but expands the asymptotic waveforms elsewhere.

By employing a procedure reminiscent of $\widetilde {\mathcal C}$, $\widetilde{\mathcal{R}}$ can be extracted to read
\bqn
\widetilde{\mathcal{R}}
= \frac{B_{\mathrm{in}}}{B_{\mathrm{out}}}e^{4i\omega x_c}
= \widetilde{\mathcal{R}}_\mathrm{BH}e^{4i\omega x_c},\nb
\eqn
which is Eq.~\eqref{TildeRForm} in the main text derived using the transfer matrix.

\section{An alternative derivation of the QNMs of the wormhole metric}\label{appC}

In this appendix, we derive the equation for wormhole quasinormal frequencies using an alternative approach based on the transfer matrix.
While the arguments follow closely that given in~\cite{agr-qnm-echoes-16}, we emphasize that the obtained equation naturally embraces both the QNMs of the original black hole and those pertaining to the wormhole leading to echoes.
The results are manifestly consistent with those obtained using different approaches.

It is not difficult to show that the corresponding transfer matrix for the entire wormhole metric is
\bqn
\mathbb{T}=\mathbb{T}^\mathrm{rhs}\cdot \mathbb{T}^\mathrm{lhs} ,
\eqn
where
\bqn
\mathbb{T}^\mathrm{rhs}=T \begin{pmatrix}e^{ i \omega x_c}&0\\0&e^{- i \omega x_c}\end{pmatrix} .
\eqn
To be specific, the transfer matrix is composed of the black hole effective potential $V_\mathrm{eff}^\mathrm{BH}(x)$ and its spatial reflection $V_\mathrm{eff}^\mathrm{BH}(-x)$.
Moreover, they are displaced to the right and left by $x_c$, which is implemented by the phase shifts in the diagonal matrix.

To evaluate the QNMs without referring to Green's function, we may enforce the outgoing wave boundary condition \eqref{master_bc2} by requiring
\bqn
\mathbb{T}_{22}=0 .\label{T22zero}
\eqn
It is straightforward to show that 
\bqn
\mathbb{T}_{22}
&=&e^{2i\omega x_c}T_{21}T_{21}^{-1}+e^{-2i\omega x_c}T_{22}T_{11}^{-1}\nb\\
&=&-T_{22}T_{11}^{-1}e^{2i\omega x_c}\left(e^{-4i\omega x_c}-\widetilde{\mathcal{R}}_\mathrm{BH}^2\right) , 
\eqn
therefore Eq.~\eqref{T22zero} implies that
\bqn
T_{22} &=& 0,\label{QNMcond1}
\eqn
or
\bqn
T_{11}^{-1} &=& 0,\label{QNMcond2}
\eqn
or
\bqn
e^{-4i\omega x_c}-\widetilde{\mathcal{R}}_\mathrm{BH}^2 =0 .\label{QNMcond3}
\eqn
Eq.~\eqref{QNMcond1} indicates that for the original black hole metric, an ``incident'' ingoing wave from $x\to -\infty$ will only produce an outgoing wave at $x\to +\infty$.
As the transfer matrix is not singular,  Eq.~\eqref{QNMcond2} is equivalent to Eq.~\eqref{QNMcond1} by exchanging the ingoing and outgoing waves.
Therefore, the original black hole's QNMs are valid solutions of Eq.~\eqref{T22zero}.
Besides, Eq.~\eqref{QNMcond3} leads to a normal spectrum associated with the echoes, as it is identical to the denominator of Eq.~\eqref{TildeCForm}.

\bibliographystyle{h-physrev}
\bibliography{references_qian}

\end{document}